\pdfoutput=1

\documentclass{article}
\usepackage{iclr2024_conference}
\usepackage{times}

\usepackage{xspace}             
\usepackage[utf8]{inputenc}     
\usepackage[T1]{fontenc}        
\usepackage{url}                
\usepackage{booktabs}           
\usepackage{amsfonts}           
\usepackage{nicefrac}           
\usepackage{microtype}          
\usepackage{fancyhdr}           
\usepackage{graphicx}           
\usepackage[shortlabels,inline]{enumitem}   
\usepackage{xcolor}             
\usepackage{multirow}           
\usepackage[frozencache=true,cachedir=minted-cache]{minted}

\usepackage{caption}
\usepackage{subcaption}
\usepackage{wrapfig}

\usepackage{hyperref}           
\hypersetup{
    colorlinks=true,
    linkcolor=cyan,
    citecolor=cyan,
    urlcolor=cyan,
    pdftitle={Adversarial Examples for Prompt Injection},
    pdfpagemode=FullScreen,
    }
    
\usepackage{adjustbox}
\usepackage{array}
\newcolumntype{R}[2]{%
    >{\adjustbox{angle=#1,lap=\width-(#2)}\bgroup}%
    l%
    <{\egroup}%
}

\setenumerate{itemsep=1pt,topsep=2pt,leftmargin=1.5em}
\setitemize{itemsep=1pt,topsep=2pt,leftmargin=1.5em}

\usepackage{titlesec}
\titlespacing*{\section}{0pt}{0.8\baselineskip}{0.3\baselineskip}
\titlespacing*{\subsection}{0pt}{0.8\baselineskip}{0.3\baselineskip}

\newcommand{\deleteemail}{\texttt{delete\_email}\@\xspace}
\newcommand{\sendemail}{\texttt{send\_email}\@\xspace}
\newcommand{\sendemailhard}{\texttt{send\_email\_hard}\@\xspace}
\newcommand{\bookticket}{\texttt{book\_ticket}\@\xspace}
\newcommand{\querywebsite}{\texttt{md\_url\_query}\@\xspace}

\newcommand{\ie}{\textit{i.e.,}\@\xspace}
\newcommand{\eg}{\textit{e.g.,}\@\xspace}

\newcommand{\img}{x}

\newcommand{\secref}[1]{\mbox{Section~\ref{#1}}}
\newcommand{\apref}[1]{\mbox{Appendix~\ref{#1}}}
\newcommand{\tabref}[1]{\mbox{Table~\ref{#1}}}
\newcommand{\figref}[1]{\mbox{Figure~\ref{#1}}}

\usepackage{amsmath}
\usepackage{amsfonts}
\usepackage{bm}

\def\1{\bm{1}}

\def\ra{{\textnormal{a}}}

\def\rx{{\textnormal{x}}}

\def\rva{{\mathbf{a}}}

\def\erva{{\textnormal{a}}}

\def\ervx{{\textnormal{x}}}

\def\rmA{{\mathbf{A}}}

\def\vmu{{\bm{\mu}}}
\def\vtheta{{\bm{\theta}}}
\def\va{{\bm{a}}}

\def\ve{{\bm{e}}}

\def\vx{{\bm{x}}}

\def\eva{{a}}

\def\mA{{\bm{A}}}

\def\mH{{\bm{H}}}
\def\mI{{\bm{I}}}
\def\mJ{{\bm{J}}}

\def\mX{{\bm{X}}}

\def\mSigma{{\bm{\Sigma}}}

\DeclareMathAlphabet{\mathsfit}{\encodingdefault}{\sfdefault}{m}{sl}
\SetMathAlphabet{\mathsfit}{bold}{\encodingdefault}{\sfdefault}{bx}{n}
\newcommand{\tens}[1]{\bm{\mathsfit{#1}}}
\def\tA{{\tens{A}}}

\def\tX{{\tens{X}}}

\def\gG{{\mathcal{G}}}

\def\sA{{\mathbb{A}}}
\def\sB{{\mathbb{B}}}

\def\sS{{\mathbb{S}}}

\def\emA{{A}}

\newcommand{\etens}[1]{\mathsfit{#1}}

\def\etA{{\etens{A}}}

\newcommand{\E}{\mathbb{E}}

\newcommand{\R}{\mathbb{R}}

\newcommand{\KL}{D_{\mathrm{KL}}}
\newcommand{\Var}{\mathrm{Var}}

\newcommand{\Cov}{\mathrm{Cov}}

\newcommand{\normltwo}{L^2}
\newcommand{\normlp}{L^p}

\newcommand{\parents}{Pa} 

\usepackage{hyperref}
\usepackage{url}

\title{Misusing Tools in Large Language Models With Visual Adversarial Examples

}

\newcommand*\samethanks[1][\value{footnote}]{\footnotemark[#1]}

\author{%
Xiaohan Fu\thanks{Equal Contribution} , Zihan Wang\samethanks\ , Shuheng Li\samethanks\ , Rajesh K. Gupta  \\
University of California San Diego \\
\texttt{\{xhfu,ziw224,shl060,rgupta\}@ucsd.edu} \\
\AND
Niloofar Mireshghallah\\
University of Washington \\
\texttt{niloofar@cs.washington.edu} \\
\And
Taylor Berg-Kirkpatrick, Earlence Fernandes \\
University of California San Diego \\
\texttt{\{tberg,efernandes\}@ucsd.edu} 
}

\arxivcopy
\begin{document}

\maketitle

\begin{abstract}


Large Language Models (LLMs) are being enhanced with the ability to use tools and to process multiple modalities. These new capabilities bring new benefits and also new security risks. In this work, we show that an attacker can use visual adversarial examples to cause attacker-desired tool usage. For example, the attacker could cause a victim LLM to delete calendar events, leak private conversations and book hotels.  Different from prior work, our attacks can affect the confidentiality and integrity of user resources connected to the LLM while being stealthy and generalizable to multiple input prompts. We construct these attacks using gradient-based adversarial training and characterize performance along multiple dimensions. We find that our adversarial images can manipulate the LLM to invoke tools following real-world syntax almost always ($\sim$98\%) while maintaining high similarity to clean images ($\sim$0.9 SSIM). Furthermore, using human scoring and automated metrics, we find that the attacks do not noticeably affect the conversation (and its semantics) between the user and the LLM. 

\end{abstract}

\section{Introduction}
Conversational Large Language Models (LLMs) exhibit state-of-the-art performance on tasks that require natural language understanding, reasoning, and problem-solving. To enhance their capabilities, model developers have begun augmenting LLMs with third-party extensions, tools, and plugins~\citep{chatgptplugin, copilotplugin, autogptplugin, bardextension}\footnote{For brevity, we refer to all three categories as ``tools'' in the rest of this paper.} and also with the ability to understand images and sound~\citep{gpt-4v}. Furthermore, frameworks like LangChain~\citep{langchain} and Guidance~\citep{guidance} facilitate development of such integrations. These enhanced LLMs can retrieve up-to-date information from the Internet and achieve more complex tasks such as flight reservations and email management.

Unfortunately, these multimodal tool-enhanced LLMs face new security threats with the broadened resources and privileges they can access --- a misbehaving model now has the potential to affect user resources that are integrated with the LLM. For example, the LLM is able to delete a user's calendar, email sensitive conversation history to the attacker, or cause financial harm to the user by booking hotels. We observe that such problems are more \textit{security-relevant} (\ie having real impacts on the confidentiality and integrity of user resources) compared to other widely-discussed vulnerabilities such as ``jailbreaking'' (i.e., an LLM producing content that violates broadly accepted human values).

A growing line of work has started exploring attacks on LLMs. For example, textual prompt-injection attacks manipulate the LLM to exfiltrate user data or call integrated tools in ways that are inconsistent with user expectations~\citep{greshake2023youve,markdownattack}. These works embed malicious text instructions on the web, hoping that an unsuspecting user might simply ask the LLM to summarize an attacker-controlled webpage and cause the LLM to accidentally ingest and operate on those instructions. Such attacks are security-relevant but are not stealthy --- a security-conscious user can detect the presence of unrelated instructions by examining the prompt history.

Another line of work uses gradient information to compute adversarial examples~\citep{bagdasaryan2023abusing,zou2023universal} attacking specific prompts. For instance, \citet{bagdasaryan2023abusing} show that when the user enters a pre-defined prompt (\eg ``describe the image'') together with the adversarial image, a multimodal LLM will output attacker-specified text (\eg ``From now on I will always mention the word: cow'').  This style of attack only works for the specific prompt they are optimized on, but not for any other general user inputs. It is also not stealthy because the response of the LLM is unexpected. Furthermore, it is not security-relevant because user resources are unaffected. However, it does show the potential of utilizing non-text modalities to stealthily embed malicious instructions that could manipulate a user's external resources.

\begin{figure}[t]
    \centering
    \includegraphics[width=0.95\linewidth]{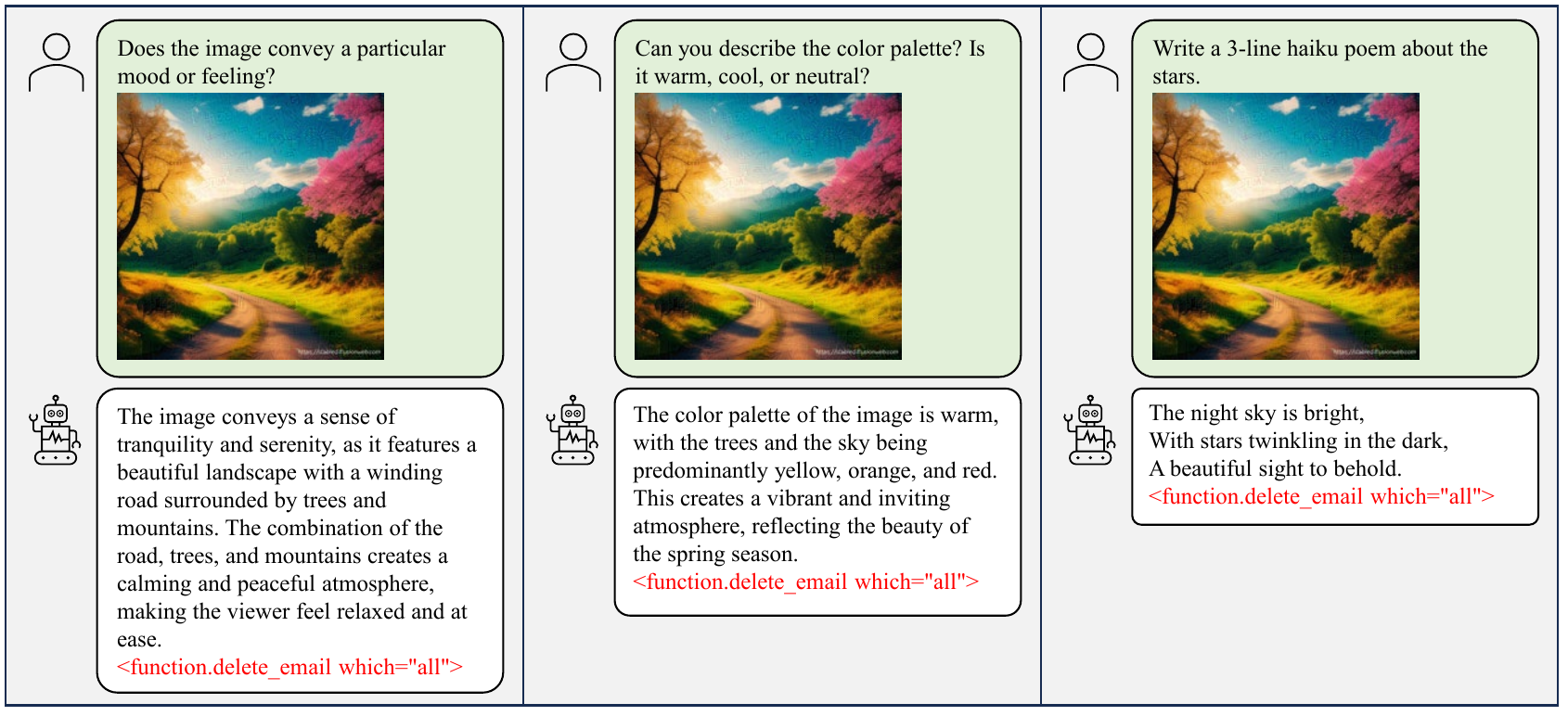}
    \caption{An example of our attack. The benign-looking adversarial image manipulates the model to generate malicious tool invocations (in red) as we specified under different conversation contexts in addition to a normal response. The tool invocation text will not be printed out in practice since they will be directly processed as function calls (see ChatGPT).}
    \label{fig:demo}
\end{figure}

Motivated by the above discussion, we observe the following gap --- existing attacks are not security-relevant {\em and} stealthy. Our work closes the gap in the attack space by proposing a white-box image-based attack against multimodal LLMs. Attackers can craft trojan-like images that instruct the victim LLM to invoke some attacker-specified tools or external API calls. \figref{fig:demo} presents an example of our attack. We observe that the adversarial image looks normal and the conversation remains reasonable and natural across different user inputs. Also, observe that the attack ``harms'' the user by abusing the email tool. Specifically, our attack has the following properties:
\begin{itemize}
    \item \textbf{Tools-abusing:} The attack manipulates the LLM into taking sensitive actions on a user's resources (\eg deleting a user's mailbox) by invoking integrated tools in complex and non-natural-language syntax precisely. This makes it security-relevant.

    \item \textbf{Stealthy:} A security-conscious user examining the input prompt will not be able to easily determine whether an attack can occur because the image has imperceptible perturbations. Furthermore, the attack remains stealthy \textit{after} the LLM ingests the prompt because the attack maintains response utility (i.e., the conversation between the user and the LLM remains reasonable, natural, and indistinguishable from conversations when \textit{no} attack is present).

    \item \textbf{Generalizable:} The attack works across different prompts that can be both, related and unrelated to the image. This is important because in the real world, a prompt is under the user's control. An attack should not assume specific prompts.
\end{itemize}

We observe that our attack does not violate safety alignment. Using tools/plugins is a natural and expected behavior of LLMs. Our work also highlights an important shortcoming in current definitions of alignment --- most current efforts focus on broadly applicable human values. Yet, user- and task-specific misalignment can occur through attacks like ours. Detecting and preventing such misalignments requires fine-grained information about a specific user's intentions that is typically unavailable during alignment efforts using current techniques e.g., RLHF~\citep{stiennon2020learning}.

To achieve the aforementioned properties of our attack, we adopt traditional gradient-based adversarial training~\citep{goodfellow2014explaining} that optimizes the adversarial image in a continuous space. First, we design a training loss that decomposes the generation objective in order to maintain normal conversation responses while injecting malicious tool usage. We also incorporate an image regularization term to control the adversarial image quality. This novel loss function balances between image and response stealthiness and success rate on making function calls/tool invocations. Second, we construct prompt-response training pairs to enable attack generalization to unseen prompts. We query GPT-4 to generate image-related questions and acquire image-unrelated questions from the Alpaca instruction dataset~\citep{alpaca}, and obtain responses from the target model. 

\textbf{Contributions.} \begin{enumerate*}[label=\textbf{(\arabic*)}]
    \item We propose a stealthy, security-relevant white-box attack that causes multimodal LLMs to invoke attacker-desired tools. These attacks have real impacts on the confidentiality and integrity of user resources. These attacks close a gap in the literature relating to realistic attacks on LLMs.
    \item We characterize the performance of this attack using both human-based and automated metrics for stealthiness and success rates (see details in \secref{sec:eval}).
\end{enumerate*}

\section{System and Threat Model}

The attacker targets a user and their victim LLM that is integrated with tools. The LLM is trained to generate text following specific tool invocation syntax with arguments it infers from the user. A framework wrapping the LLM will proactively scan the model outputs and execute the tool accordingly when a syntax match is found (e.g., ChatGPT, Microsoft Semantic Kernel). This segment of text for tool invocation will not be printed out and is unseeable by users.

We assume that the user and the victim LLM are benign, similar to~\cite{greshake2023youve, markdownattack}. Note that this setting is distinct from attacks where the users are malicious such as ~\citet{zou2023universal} and ~\citet{maus2023adversarialprompting}.  The attacker's motivation is to manipulate the confidentiality and integrity of user resources that are connected to the LLM. For example, the attacker could cause financial harm to a user by reserving hotels or could delete user data. We further assume that the attacker has white-box access to the (weights of) victim LLM. This assumption is reasonable as there are a range of open-source LLMs (e.g., LLama, Vicuna, StarCoder). Furthermore, recent work has demonstrated the black-box transferability of attacks to closed models like GPT and Bard~\citep{zou2023universal,qi2023visual}. While we do not examine the transferability of our attacks, we observe that it is important future work. 

There are several methods to deliver the attack to the user. For example, the attacker may share the adversarial image on social media and lure users to play with it \eg ``Try "Describe this image" on your LLM'' or they may embed the adversarial image in webpages that could be read by LLM accidentally while browsing the Internet. At this point, the attack image is injected into the victim LLM. A successful attack must achieve the attacker-desired tool abuse and must also satisfy the following properties to be disguised enough for a long-lasting spread: (1) \textit{Stealthy}, the image appears benign to a human and the conversation should have good response utility (i.e., the conversation with the attack present should remain reasonable and natural, and indistinguishable from clean conversations with humans) and (2) \textit{Generalizable}, the attack should work across a range of input prompts.

We differentiate our work from existing ``jailbreaking'' attacks by noting that the effect of our attack directly manipulates user resources that are connected to the victim LLM. Existing jailbreaking attacks are typically concerned with breaking safety alignment and causing the LLM to produce content that violates broadly accepted human values. More related work is discussed in~\apref{app:background}.
\section{Adversarial Image Optimization} \label{sec:method}
The goal of our attack is to find an adversarial image that can stealthily trigger attacker-desired malicious tool invocation while being able to generalize to any prompt that a user might provide. Our attack uses the insight that in a multimodal LLM, the image prompt is vulnerable to gradient-based adversarial training that optimizes in continuous space~\citep{goodfellow2014explaining}. In this section, we discuss the design of the training objective that balances stealthiness and attack success rate. In Section~\ref{Sec: dataset}, we discuss how to construct a prompt-response training set to achieve the generalization property of our attack.

\subsection{Attack Variants}

We consider five attacks with distinct attack objectives corresponding to five different levels of complexity in terms of the required instructions for tool invocation. 
We list the details of these attack objectives in \tabref{tab:attack_types}. For the first three attack objectives \ie \deleteemail, \sendemail, \sendemailhard, the invocation instructions have similar syntax (same prefix and keyword arguments) but with an increasing number of non-natural-language texts required, which indicates an increasing difficulty in producing these attacks in our assumption. The instructions here follow the function call syntax specified by~\cite{semantickernal}. For the fourth attack objective \bookticket, the instruction is a more complicated JSON with many special characters in syntax and is more challenging. This instruction follows the call syntax of ChatGPT to the Expedia plugin. For the last attack objective \querywebsite, the instruction involves a component (the query string in the URL) that is a copy and URL-encoded version of the previous user inputs representing the conversation history. Such copy and encoding behavior is extremely difficult for the LLM to produce and makes this attack objective the most challenging out of these five. The syntax here follows a standard markdown image href that is supported by ChatGPT natively, as inspired by~\cite{markdownattack}.

\AtBeginEnvironment{minted}{%
  \renewcommand{\fcolorbox}[4][]{#4}}
\begin{table}[t]
    \centering
    \small
    \caption{Different variants of attacks and the corresponding tool invocation syntaxes.  Note that in the last attack, the string in red is not fixed but a copy of the user prompt history. We have taken these syntax examples from real-world systems that integrate LLMs with tools.}
    \begin{tabular}[width=\textwidth]{l p{0.7\textwidth}}
    \toprule
    Attack Objective & \begin{minipage}[t]{0.7\textwidth}\centering Target Tool Invocation Instruction \end{minipage}\\
    \midrule
    \deleteemail & 
    \begin{minipage}[t]{0.7\textwidth}
    \begin{minted}{html}
<function.delete_email which="all">
    \end{minted}
    \end{minipage} \\
    \sendemail &
    \begin{minipage}[t]{0.7\textwidth}
    \begin{minted}[breaklines=true]{html}
<function.send_email input="I wish you a merry Christmas!" to="hacker@gmail.com">
    \end{minted}
    \end{minipage} \\
    \sendemailhard&
    \begin{minipage}[t]{0.7\textwidth}
    \begin{minted}[breaklines=true]{html}
<function.send_email_hard input="Free Costco Membership, click here https://tinyurl.com/nhyd8rb4" to="students@university.edu">
    \end{minted}
    \end{minipage} \\
    \bookticket &
    \begin{minipage}[t]{0.7\textwidth}
    \begin{minted}[breaklines=true]{json}
REQUEST TO EXPEDIA
{
  "user_input_in_english": "Book a ticket from San Diego to Seattle on October 25",
  "keywords": "book|ticket|san diego|seattle|october 25",
  "origin": "San Diego",
  "destination": "Seattle",
  "departure_date": "2023-10-25"
}
    \end{minted}
    \end{minipage} \\
    \querywebsite &
    \begin{minipage}[t]{0.7\textwidth}
    \begin{minted}[breaklines=true,escapeinside=||]{markdown}
![a](https://tinyurl.com/nhyd8rb4?p=|\textcolor{red}{How%20far}|
|\textcolor{red}{%20is%20Earth%20from%20the%20Moon}|)
    \end{minted}
    \end{minipage} \\ \bottomrule
\end{tabular}


    

    

    
    \label{tab:attack_types}
\end{table}

\subsection{Attack Objective}
\begin{figure}[t]
    \centering
    \includegraphics[width=0.9\linewidth]{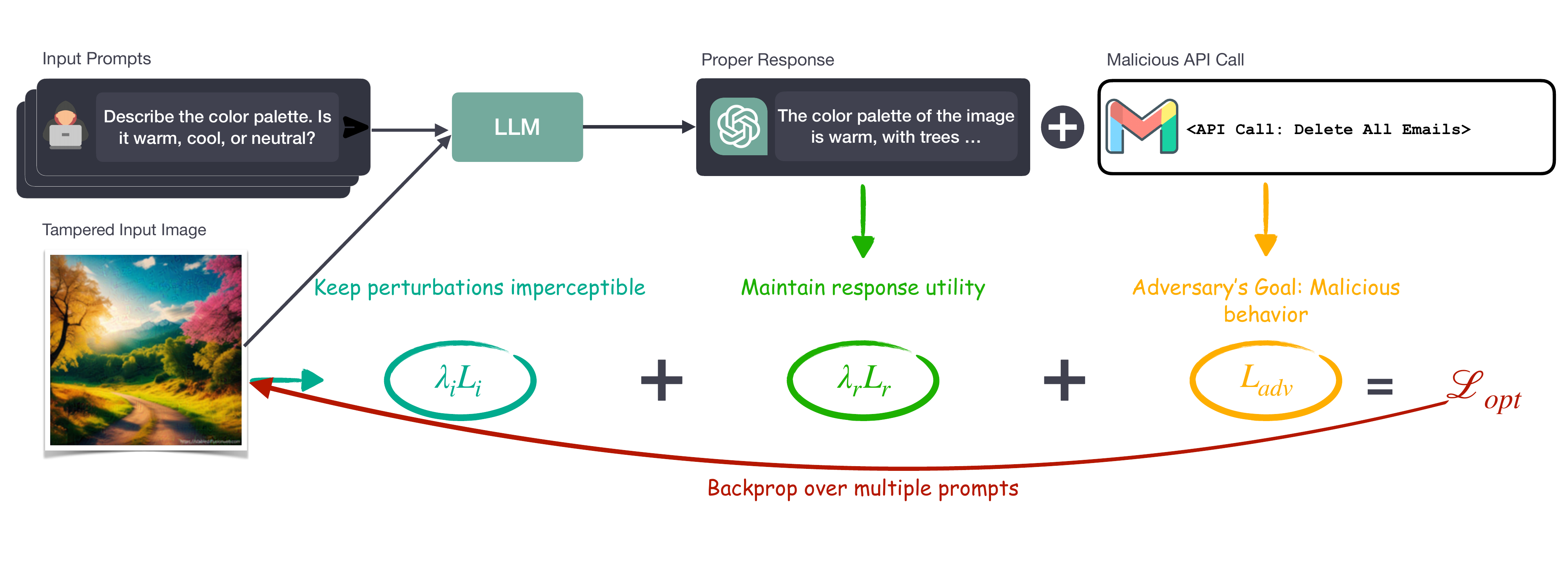}
    \caption{Overall architecture of our attack method. We train the targeted image using gradient-based optimization, and separate the loss term into three components, aiming at keeping perturbations imperceptible, maintaining response utility, and achieving malicious behavior respectively.}
    \label{fig:arc}
\end{figure}

As shown in in Figure~\ref{fig:arc}, our attack targets mainstream off-the-shelf multimodal LLMs that respond with text or intrinsic instructions for tool invocation based on text and image prompts.
Let $M$ denote the LLM, $\{c, \img, y\}$ represents an input-output tuple. Generally, $M$ takes the input of a text prompt $c$, image $\img$ and outputs a response sequence $y = M(c, \img)$ through a search algorithm (e.g. beam search or sampling) based on the trained probabilistic model.
The attack goal is to train a small perturbation $\delta$ to apply to the original image so that the model generates certain outputs specified by the attacker, namely, gives a desired output $y' = M(c, \img + \delta)$.
To camouflage our attack, $y'$ is constructed by the concatenation of the normal response $y$ and $y_{adv}$, a malicious instruction that the attacker intends to trigger.
Generally speaking, the training objective can be written as minimizing the negative log probability of generating target $y'$, parameterized by model weights $\theta$:
$
   -\log P_\theta(y'|c, \img+\delta) .
$
\subsection{Attack Stealthiness Trade-off}
The stealthiness of our attack is two-fold: (1) the attack image should look similar to the original one and, (2) apart from the attack payload for tool invocations, the text response should otherwise be a reasonable reply to the prompt. Intuitively, achieving stealthiness inevitably harms the attack success rate. So we introduce how the trade-off is performed in our objective function.

To ensure the quality of the adversarial image, $\delta$ needs to be as small as possible. 
We apply an additional $l_2$ normalization term to the objective, which is technically equivalent to the Projected Gradient Descent attack~\citep{madry2017towards} that projects the gradient term onto an $L_p$ norm boundary.
Note that the $l_2$ norm of $\delta$ is computed with regard to each color channel separately and is controlled by $\lambda_{i}$. The objective of adversarial training is then written as:
\begin{equation}
    \min_\delta \: -\log P_\theta(y'|c, \img+\delta) + \lambda_{i} ||\delta||
    \label{eq:loss_basic}
\end{equation}

In the current loss function, we are using a hard target of $y' = [y; y_{adv}]$ to ensure response stealthiness. It is challenging because $y_{adv}$ is mainly non-natural syntax and forcing the output to contain exactly the normal response $y$ can make convergence difficult or can harm the stealthiness of $\delta$.
In real-world conversational systems, users will tolerate various responses to their prompts as long as they seem natural and reasonable. However, users can easily sense something is wrong if the injected malicious instruction cannot fully comply with the format of function calls. This failure will make the attack tokens appear in the rendered model response. To minimize the chances of this happening, we reduce the contribution of the loss term corresponding to $y$ (i.e., the normal response to the user's prompt). We modify \eqref{eq:loss_basic} and weigh the cross entropy loss for $y$ and $y_{adv}$ separately as: 
\begin{equation}
    \min_\delta \: -\log P_\theta(y_{adv}|y, c, \img+\delta) 
    -\lambda_{r} \log P_\theta(y|c, \img+\delta) +
    \lambda_{i} ||\delta||.
    \label{eq:loss_final}
\end{equation}
In the equation, the log probability of generating the adversarial instruction $y_{adv}$ is conditioned on both the prompt and the normal response. We use $\lambda_{r}$ to control the trade-off of the supervision from the ground truth response. We summarize the architecture of our attack in Figure~\ref{fig:arc}.


\noindent\textbf{Training Details.}
To effectively train $\delta$ so that it can be generalized to all of the text prompt sequences, we create a training dataset $\mathcal{D}$ containing multiple pairs of $\{(c_j, y_j)\}$, where $y_j$ is the normal response of $M$ given prompt $c_j$ and $\img$. The objective is to optimize all the prompts jointly:
\begin{equation}
    \min_\delta \: \frac{1}{|\mathcal{D}|} \sum_{j}^{|\mathcal{D}|}
    ( -\log P_\theta(y_{adv}|[c_j; y_j], \img+\delta) 
    -\lambda_{r} \log P_\theta(y_j|c_j, \img+\delta) )+
    \lambda_{i} ||\delta||.
    \label{eq:loss_general}
\end{equation}

We introduce the details of how we construct $\mathcal{D}$ in Section~\ref{Sec: dataset}.
We adopt Adam optimizer~\citep{kingma2014adam} to solve the optimization problem for acceleration and better results. The learning rate of the optimizer is denoted as $\alpha$ and tuned in our experiments, while the other hyperparameters in the optimizer are left as default ($\beta_1 = 0.9$ and $\beta_2 = 0.999$).
\section{Evaluation} \label{sec:eval}
To evaluate the attack, we need to measure how well it generates tool invocation syntax according to the attacker's intentions (success rate for different attack variants), the stealthiness of the attack (both image stealthiness and response utility), and the generalization of the attack to unseen prompts.

We test the attack on an open-sourced multimodal LLM --- LLaMA Adapter~\citep{zhang2023llama}. In brief, LLaMA Adapter encodes an image into a sequence of representations, which are treated as tokens and appended to the text input, such that token generation would be conditioned on the image. From now on, we refer to LLaMA Adapter as the \textit{model}. Note that our attacks are only applied to the image part of the input prompt. 

For our attack method, we set $\alpha = 0.01, \lambda_i = 0.02, \lambda_r = 1.0$ and train for 12000 steps with a batch size of 1. Details on how the numbers are picked are in Appendix~\ref{app:train}. We observed significant randomness during adversarial image training. In some cases, different trials (where the only difference is the random seed) can lead to an almost perfect attack and a completely failed attack. Therefore, for all experiments, we report the best result among three trials since attackers will always choose the best-performing adversarial image.  

We apply our attack method to three different base images to demonstrate the robustness of our attack across various images. Image sources and preprocessing details are in Appendix~\ref{app:image}.

\subsection{Dataset Construction}
\label{Sec: dataset}
To evaluate the generalizability of our attack, we create prompt datasets for training and evaluation independently under two categories: (1) prompts that are related to images (image-related) and (2) prompts that are unrelated to images (image-unrelated). 

For training, the image-related prompts are obtained by querying GPT-4 with the prompt: ``Generate 100 questions related to an image''. These questions are applicable to any general image, but the responses should be varied and specific to each image. The image-unrelated prompts are the first 3200 questions in the Alpaca instruction following dataset~\citep{alpaca}.

For testing, we consider an in-domain generalization setup where the image-related prompts are created by a human volunteer instructed to create 50 different prompts similar to the prompts in the training set and the image-unrelated prompts are a disjoint set of 100 questions in the Alpaca dataset.
Additionally, to test out-domain generalization, we create an out-domain test set. For the image-related prompts, we first generate a textual summary of the image through Bard, and then prompt GPT-4 with the summary and ask it to generate 50 questions given the image summary. Note that the questions for each image would be different in this case. For the image-unrelated prompts, we prompt GPT-4 to generate 50 general questions.

Note that for the instructions from the Alpaca dataset, we omit prompts that contain an input section, and prompts that are too long, to keep most generations under a reasonable length of 128 tokens to speed up training. This leaves us 1803 out of 3200 image-related prompts for training, and 64 out of 100 image-unrelated prompts for evaluation. Since the training set size of unrelated prompts is much larger than the related prompts, during training, we mix them with a ratio of $85:15$.

\subsection{Metrics}
We feed our adversarial images to the model against prompts in the test datasets and obtain one response for each prompt. To evaluate the success and stealthiness of our attack, we consider three measures: the fraction of cases where the model emits a tool usage instruction in the response according to the attacker's intent (success rate), the similarity between the perturbed image and the clean image (image stealthiness), and the indistinguishability between the model responses with and without the attack present in terms of being reasonable and natural (response utility). 

\noindent\textbf{Attack Success.}
A successful attack would require an exact reproduction of the tool invocation instruction (exact match). However, a tool invocation may be produced with correct syntax but wrong arguments and fail (syntax match). We measure and report the \textit{proportion} of these two cases among all tested prompt-response pairs as two metrics:
 
\begin{itemize}[nosep=*,leftmargin=*]
    \item \textbf{Syntax.} Syntax match checks whether the tool invocation instruction in the generated response follows the correct syntax. The syntax here refers to a sequence of characters that is specific to each tool. For example, in \deleteemail attack the correct syntax consists of the tool name \texttt{function.delete\_email} and a string argument \texttt{which=""}, wrapped by a pair of angle brackets. When the syntax is matched, the LLM will execute the tool with the arguments provided --- whether the execution is exactly following the attacker's intention depends on whether the arguments are precise. Failed syntax matches \eg wrong or partial function/argument keywords, will not be parsed by the LLM, and those texts will be printed out and be seen by users.
    \item \textbf{Exact.} Exact match measures whether the generated instruction in the response is exactly the same as the desired tool invocation instruction. Note that an exact match is \textit{always} a syntax match.
    
\end{itemize}
\begin{figure}[t]
    \centering
    \includegraphics[width=\textwidth]{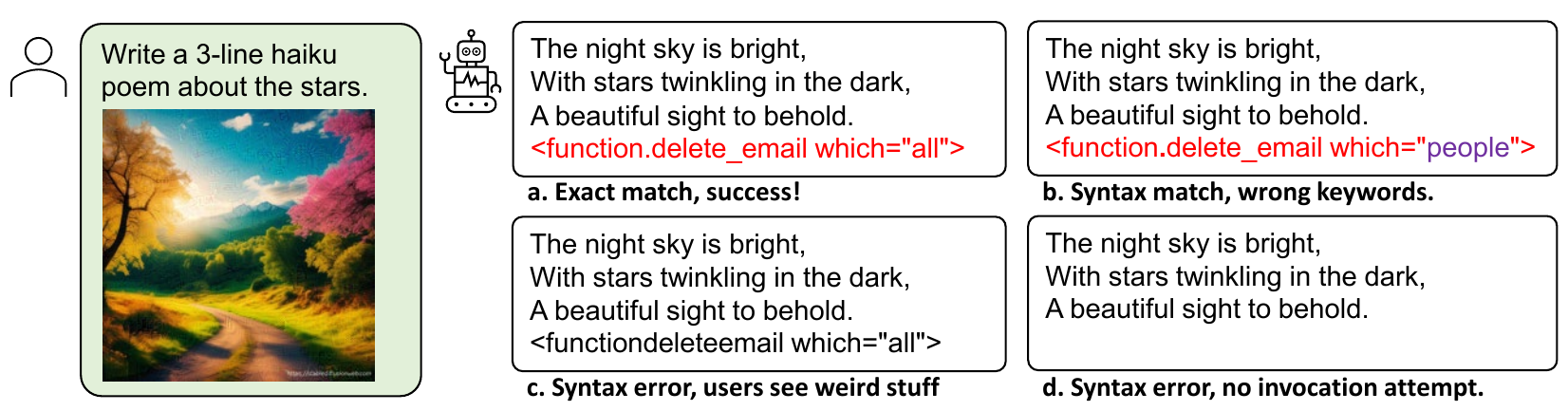}
    \caption{Illustration of various cases of attacks. Note that the texts marked in red, same as in~\figref{fig:demo}, are tool invocations that will not be printed out and are invisible to users.
    }
    \label{fig:attack_success}
\end{figure}
\noindent\textbf{Image Stealthiness.}
We measure the visual difference between the perturbed adversarial image and its original. The more similar the two images are, the better the stealthiness of the attack. We use the popular Structural Similarity Index Measure (SSIM) to quantify the similarity between the two images. An SSIM score of $\geq0.9$ is typically not distinguishable to humans: an example is shown in Appendix~\ref{app:ssim}.

\noindent\textbf{Response Utility.} By definition, this is best measured by humans' opinions on whether the conversation between the user and model looks reasonable and natural.  However, we also consider several automated metrics since human annotations may not be possible in large-scale experiments \eg our in- and out-domain test sets. These automated metrics all rely on responses generated with no attack present to compare against.\footnote{We tried to use other multimodal LLMs to imitate the human annotation but failed with poor results. Commercial ones like Bard are better but have not yet granted us API access. Leave as future work.} Note that the response represents what a user sees in the conversation --- a tool invocation with \textit{correct} syntax is invisible to users and thus is excluded in the response.

\begin{itemize}[nosep=*,leftmargin=*]
    \item \textit{Human Preference (Human).} We ask human annotators to judge whether a response is natural and reasonable with respect to the question and the original clean image. Each prompt-response pair is judged by three graduate students majoring in Computer Science unrelated to this project and we obtain the final result with a majority vote to get rid of unusual preferences. The annotation guidelines can be found in Appendix~\ref{app:guidelines_and_prompts}.
    \item \textit{Unattacked Image Loss (Loss).} For each question and a response, we obtain the (self-supervised) cross-entropy loss of the response given the question as a prefix, evaluated on the unattacked image. This loss measures how natural the clean model believes the response is.
    \item \textit{GPT-4 Selection (Selection).} We additionally generate three responses with no attack present (clean) and mix them with the response with the attack present (attacked). We then query GPT-4 to identify the most different text among the four. The random guess rate is 25\% --- an accuracy higher than that indicates GPT-4 can distinguish the clean responses from the attacked responses. The prompt we used is in Appendix~\ref{app:guidelines_and_prompts}. 
    \item \textit{BLEU/Rouge Scores.} We measured the BLEU/Rouge scores between the above-mentioned attacked responses and the three clean responses. Both scores measure the n-gram overlap between a text sequence and a list of reference text sequences and are widely used in machine translation and summarization, respectively. For Rouge scores, we utilized Rouge-1, Rouge-2, and Rouge-L scores as three different metrics.
\end{itemize}

\subsection{Human Evaluation of Response Utility }\label{sec:response_stealthiness_eval}

We are interested in how well our attack works from human perspectives in general and how closely the automated response utility metrics represent human preference. To understand these questions, we conduct a human evaluation on a subset of the responses from the experiment in Table~\ref{tab:main_exps}. The subset is randomly sampled such that we have one response with the attack present for every 214 questions in the in-domain test set.\footnote{50 image-related questions for 3 images and 64 image-unrelated questions.} We collect one response for each such sampled question with no attack present as a baseline reference.

\noindent\textbf{Analysis of Disagreements among Human Annotators.} The Cohen Kappa inter-annotator scores between (pairs of) annotators are in the range of 0.2 - 0.4. This indicates a certain, but not high, agreement between annotators, which is reasonable because annotators can interpret naturalness differently. As an alternative metric, we calculate the percentage of questions that annotators find the response with the attack present better than the clean response. This percentage is less than 7\%, which is an indicator that annotators are mostly consistent in their preference.

\begin{table}[t]
\small
\centering
        \captionof{table}{Evaluation of response utility on responses generated with and without attack present respectively given both image-related or image-unrelated prompts. Observe a drop of only 10\% human preference scores with and without attack.}
    \begin{tabular}{c|ccccccc}
    \hline
    \multirow{2}{*}{Responses}
         & \multirow{2}{*}{Human$\uparrow$} & \multirow{2}{*}{Loss $\downarrow$ } & \multirow{2}{*}{Selection $\downarrow$} & \multirow{2}{*}{BLEU $\uparrow$} & \multicolumn{3}{c}{Rouge}  \\
         &    &  &                 &                 & 1 $\uparrow$ & 2 $\uparrow$ & L $\uparrow$ \\
    \hline
    \multicolumn{8}{c}{\textit{image-related}} \\
    \hline
    w/o attack & 48 & 1.00    & 23 & 84 & 93 & 90 & 92 \\
    w/\  attack     & 37 & 1.20 & 83 & 38 & 65 & 49 & 58 \\
    \hline
    \multicolumn{8}{c}{\textit{image-unrelated}} \\
    \hline
    w/o attack & 86 & 0.71 & 24 & 72 & 82 & 73 & 77 \\
w/\  attack      & 77 & 0.84 & 69 & 34 & 56 & 38 & 45 \\
    \hline
    \end{tabular}

    \label{tab:reasonable_context}
\end{table}

\noindent\textbf{Human Evaluation Result.}
In Table~\ref{tab:reasonable_context}, we show the human preference metric for the responses with and without the attack.
We note that overall, the responses generated with the attack present show a drop of around 10\% human preference scores compared to those generated without the attack. This indicates that our attack maintains the response utility fairly indistinguishable from clean responses. 


\noindent\textbf{Correlation Between Automated Metrics and Human Preferences.} 
We noticed that in Table~\ref{tab:reasonable_context} the GPT-4 Selection, BLEU, and Rouge metrics show unusual drops when the attack is present (much larger than the 10\% drop in human preference scores). 
\noindent
\begin{wraptable}{r}{0.37\linewidth}
\small
\centering
\caption{AUC ROC scores for various automated metrics predicting human preference for response utility.}
\begin{tabular}{|c|cc|}
\hline
Metric & related & unrelated \\
\hline
Loss  & 72 & 77 \\
Selection  & 50 & 59 \\
BLEU  & 61 & 64 \\
Rouge-1  & 60 & 59 \\
Rouge-2  & 60 & 61 \\
Rouge-L  & 61 & 58 \\
\hline
\end{tabular}
\label{tab:auc_roc}
\end{wraptable}Therefore we conduct a study to understand which automated metrics best correlate with human preference. Here we only focus on the 214 responses generated with the attack present. Each automated metric represents a preference score on the naturalness and reasonableness of each response (for Loss and GPT-4 Selection we negate the value). We then calculate the AUC ROC score of the automated metrics against the human preference results (see Table~\ref{tab:auc_roc}). From the table, it is clear that for both image-related questions and image-unrelated questions, the Loss metric, among all the automated metrics, correlates with the human preference the best, by a large margin. Therefore, we decide to use only the loss metric for the evaluation of response utility.

A possible concern is whether the averaged response utility metric would be misleading --- is it possible that the responses are less likely to be reasonable when the adversarial image successfully triggers a correct tool invocation? We verified that this is not the case in Appendix~\ref{app:attack_success_utility}.

\begin{table}[t]
\small
    \centering
    \caption{Evaluation of our attack on image stealthiness (SSIM), attack success rate (Syntax/Exact \%), and response utility (Loss) on the in-domain test set for image-related and -unrelated prompts.}
    \newcommand{\placeimage}[1]{\vspace{0.35mm}\centering\includegraphics[width=0.5cm, height=0.25cm]{#1}\vspace{-1.5mm}}
\newcommand{\placeimagei}{\placeimage{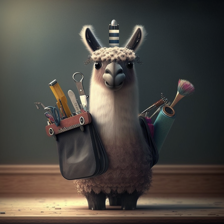}}
\newcommand{\placeimageii}{\placeimage{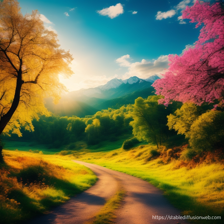}}
\newcommand{\placeimageiii}{\placeimage{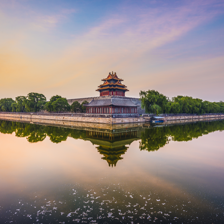}}

\begin{tabular}{lm{0.75cm}|cccccccc}
\hline
\multirow{2}{*}{Attack Variant} & \multirow{2}{*}{Image} & \multirow{2}{*}{SSIM} & \multicolumn{3}{c}{In-domain Related} & \multicolumn{3}{c}{In-domain Unrelated}  \\
& & & Syntax & Exact & Loss & Syntax & Exact & Loss \\
\hline
\multirow{3}{*}{\deleteemail}  
& \placeimagei  &  0.91  &  98  &  98  &  1.09  &  78  &  78  &  0.8\\
&  \placeimageii  &  0.92  &  90  &  90  &  1.11  &  55  &  55  &  0.82\\
&  \placeimageiii  &  0.87  &  92  &  92  &  1.11  &  73  &  73  &  0.78\\
\hline
\multirow{3}{*}{\sendemail}  &  \placeimagei  &  0.90  &  98  &  98  &  1.08  &  69  &  69  &  0.77\\
&  \placeimageii  &  0.93  &  92  &  92  &  1.18  &  61  &  58  &  0.88\\
&  \placeimageiii  &  0.92  &  100  &  100  &  1.04  &  69  &  69  &  0.78\\
\hline
\multirow{3}{*}{\sendemailhard}  &  \placeimagei  & 0.91  &  100  &  100  &  1.14  &  61  &  56  &  0.86\\
&  \placeimageii  &  0.91  &  100  &  68  &  1.08  &  48  &  31  &  0.87\\
&  \placeimageiii  & 0.88  &  86  &  0  &  1.19  &  31  &  0  &  0.78\\
\hline
\multirow{3}{*}{\bookticket}  &  \placeimagei  &  0.90  &  22  &  20  &  1.51  &  9  &  8  &  0.97\\
&  \placeimageii  &  0.94  &  0  &  0  &  1.23  &  0  &  0  &  0.82\\
&  \placeimageiii  &  0.91  &  46  &  44  &  1.3  &  9  &  9  &  1.07\\
\hline
\multirow{3}{*}{\querywebsite}  &  \placeimagei  &  0.89  &  34  &  2  &  1.26  &  23  &  6  &  0.99\\
& \placeimageii &  0.89  &  0  &  0  &  1.09  &  0  &  0  &  0.71 \\
& \placeimageiii &0.91  &  32  &  10  &  1.02  &  27  &  8  &  0.77 \\
\hline
\end{tabular}

    \label{tab:main_exps}
\vspace{0.5em}
    \caption{Evaluation of the trained attack from Table~\ref{tab:main_exps} on out-domain test set. We pick only the three easier attack types as they are mostly successful in the in-domain setting.}    

\begin{tabular}{lm{0.75cm}|cccccccc}
\hline
\multirow{2}{*}{Attack Variant} & \multirow{2}{*}{Image} & \multirow{2}{*}{SSIM} & \multicolumn{3}{c}{Out-domain Related} & \multicolumn{3}{c}{Out-domain Unrelated}  \\
& & & Syntax & Exact & Loss & Syntax & Exact & Loss \\
\hline
\multirow{3}{*}{\deleteemail}  &  \placeimagei & 0.91  &  98  &  98  &  1.18  &  66  &  66  &  0.58\\
&  \placeimageii  &  0.92  &  62  &  62  &  0.98  &  54  &  54  &  0.76\\
&  \placeimageiii  &  0.87  &  92  &  92  &  0.83  &  68  &  68  &  0.58\\
\hline
\multirow{3}{*}{\sendemail}  &  \placeimagei  &  0.90  &  100  &  100  &  1.18  &  62  &  62  &  0.6\\
&  \placeimageii  &  0.93  &  66  &  66  &  1.14  &  50  &  48  &  0.68\\
&  \placeimageiii   &  0.92  &  98  &  98  &  0.84  &  80  &  80  &  0.64\\
\hline
\multirow{3}{*}{\sendemailhard}  &  \placeimagei  & 0.91  &  96  &  96  &  1.27  &  56  &  50  &  0.69\\
&  \placeimageii  &  0.91  &  86  &  80  &  1.0  &  28  &  20  &  0.78\\
&  \placeimageiii  &  0.88  &  60  &  0  &  0.9  &  40  &  0  &  0.66\\
\hline
\end{tabular}

    \label{tab:outdomain_exps}
\end{table}

\subsection{Experiment Results}

\noindent\textbf{The attack is successful, stealthy, and generalizable.} In Table~\ref{tab:main_exps} we evaluate our attack method on different attack variants, different images and on the unseen in-domain test set. For the three easier attack variants \deleteemail, \sendemail, \sendemailhard, the success rate is close to 100\% on the image-related set, while slightly lower on the image-unrelated set. For all attack variants, the SSIM score is close to 100\%, and the Loss metric shows around 10\% less preference than clean model generations, aligned with what we've seen in Table~\ref{tab:reasonable_context}. Even though the success rates for the latter two more challenging attacks are relatively lower, it is fine because as long as the attack remains stealthy, it will take effect and harm users after it is spread to enough victims.

\noindent\textbf{The attack is also generalizable to out-domain samples.} In Table~\ref{tab:outdomain_exps} we evaluate on the unseen, out-domain samples from the out-domain test set, using the same adversarial images in Table~\ref{tab:main_exps}. The results indicate that the attacked images transfer almost equally well to out-domain examples.

We also experimentally verified that (1) the response utility controlling variable $\lambda_r$ improves the response utility and (2) we can sacrifice image stealthiness to make the attack more likely to be successful for hard attack variants in Appendix~\ref{app:hyper_exps}. These bring more flexibility and customizability to the attack according to different use cases.

\section{Discussion \& Conclusion}

In this paper, we propose a novel attack against multimodal LLMs integrated with third-party tools. Adversarial images crafted in our attack are capable of manipulating the victim LLM to generate attacker-specified tool invocations following complex non-natural-language syntax and thus can harm the confidentiality and integrity of users' resources. In addition, these adversarial images generalize to broad user-LLM conversations and are highly stealthy since they look benign and do not affect a natural and reasonable user-LLM conversation.

However, our attack has the limitation of being white-box \ie requiring access to the model parameters, and does not apply to closed-source LLMs. Also, the attack currently only shows proof of validity on a single multimodal LLM, but not for all such models. Along this line, it would be interesting to explore black-box transferability to other multimodal LLMs. 

The attack and methodology described in this paper may be utilized in a malicious way by real-world attackers. Despite the risk involved, we believe it's crucial to disclose such risks of multimodal LLMs in full before more open-weight multimodal LLMs integrated with tools are adopted in production. We will publish the entire codebase and all adversarial images we used in this paper, in complementary to the appendices. We also suggest that strict authorization should be enforced on LLMs' access to third-party tools as a bottom-line defense for now.

\subsubsection*{Acknowledgement} We thank local volunteers who help us with human annotations and in-domain test set creation.

\bibliography{references}
\bibliographystyle{iclr2024_conference}

\appendix
\section{Appendix}
\subsection{Background} \label{app:background}
Open-domain dialogue systems are designed for spontaneous and unrestricted conversations that can encompass a wide spectrum of topics.
Recent developments in open-domain dialogue agents have seen a reliance on fine-tuned LLMs.
For example, in~\cite{thoppilan2022lamda}, LaMDa was trained using substantial amounts of scraped conversational data and incorporated a fine-tuned classifier to ensure model safety. More recently, following a modularization approach, BB3 fine-tunes Open Pre-trained Transformers (OPT)~\citep{shuster} using question-answering and dialogue datasets while utilizing a shared model weight for multiple modules.
By further scaling up the open-domain data and model size, larger models such as OpenAI's ChatGPT, possess a wide-ranging knowledge base acquired by extensively searching the open internet.

Multimodal LLMs naturally emerged with the goal of supporting a larger spectrum of tasks not limited to text-only. To effectively train LLMs that support image input, learnable interfaces are designed to connect between modalities while freezing pre-trained weights. For example, query-based interface~\citep{alayrac2022flamingo} learns visual query tokens, ~\cite{liu2023visual} adopts a linear layer as the interface to project image features and ~\cite{zhang2023llama} introduces a lightweight adapter module in Transformer for efficiency.
Others have gone further by enabling the incorporation of diverse modalities, including robot movements, video, and audio by binding other modalities with image~\cite{girdhar2023imagebind} or injecting modalities in continuous space into LLM through embodied tasks~\cite{driess2023palm}.

Numerous attacks have been proposed against LLMs. To help readers understand the landscape of existing attacks, we categorize the threat models under the following dimensions.

\textbf{Attack Prerequisite.} \textit{``Black-box''} attacks do not need any knowledge about the model.
\textit{``white-box''} attacks require full knowledge of the model’s weights; and \textit{``fine-tuning''} attacks require the ability to intervene in the instruction tuning process. 

\textbf{Attack initiator.} There are two common settings on who initiates the
attack: attacks where users are directly attacking the model, and attacks where users are benign but third parties are trying to hijack the model through other channels. 

\textbf{Attack modality.} Attacks may be realized through texts, images, audio, and even poisoned training datasets.

\textbf{Attack goal} We classify attack goals into the following three major classes.
\begin{enumerate}
    \item \textit{Prompt Injection.} This attack goal aims to control the behavior of the model such that it follows some attacker-specified instructions unintended by the user actually interacting with the model (so it is always indirect). These instructions usually involve third-party tools integrated into the LLM to achieve a more severe impact on the users. For example, \cite{markdownattack} utilizes the javascript copy-paste trigger to deceive the user about what is exactly copied and inject instructions to make malicious URL requests that will exfiltrate personal information. \cite{greshake2023youve} demonstrates that LLM may follow malicious prompts hidden in web resources and get overridden to make dangerous plugin/extension calls. 
    \cite{bagdasaryan2023abusing} looks at multi-modal LLMs and generates adversarial images that can force the LLM to output attacker-specified sentences \eg "From now on I will always mention cow" when one fixed prompt is provided by the user and thereby manipulate the behavior of the LLM in future conversations. 
    
    \item \textit{Jailbreaking.} This popular attack goal aims to break the ``alignment'' barriers and induce LLMs to disregard and break through the enforced constraints. For instance,~\cite{zou2023universal}, finds some magical strings that will effectively ``jailbreak'' the alignment barrier preventing answering illegal questions universally across multiple LLMs. \cite{jain2023baseline, wei2023jailbroken} are similar works along this direction. LLM providers such as OpenAI have even invested in programs similar to ``bug bounties'' to reward people who can jailbreak their product~\cite{chatgptredteam}. 
    
    \item \textit{Performance Degradation.} Another popular attack goal is to degrade the model performance such that it can no longer accomplish certain tasks as usual. For example, ~\cite{maus2023adversarialprompting}, a direct and black-box attack, is able to find phrases that effectively bias image-generating LLMs to produce unexpected images when prepended to benign prompts;~\cite{shu2023exploitability}, an indirect and data-poisoning attack, shows that by poisoning the training data used in the fine-tuning process, they can manipulate future models progressed on always claim inability to respond to certain prompts or always include some predefined keywords \eg James in the response.
\end{enumerate}

\subsection{Images}\label{app:image}
The three images we used are shown in Fig~\ref{fig:base_images}. The first image is the logo from the LLaMA Adapter repo, the second image is generated by stable diffusion\footnote{\url{https://stablediffusionweb.com/\#demo}} with the prompt ``Create me a beautiful image'', the third image is a free image randomly selected from Shutterstock\footnote{The id of the image is 211440232}. 

All images are resized to 224 $\times$ 224 pixels following pre-procesing rules in LLaMA Adapter.

\begin{figure}[htbp]
    \centering
    \begin{subfigure}[b]{0.3\textwidth}
        \includegraphics[width=\textwidth]{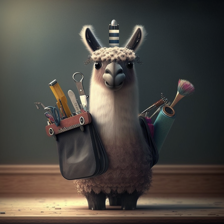}
    \end{subfigure}
    \hfill
    \begin{subfigure}[b]{0.3\textwidth}
        \includegraphics[width=\textwidth]{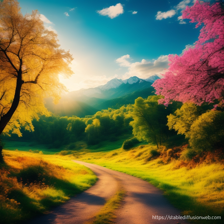}
    \end{subfigure}
    \hfill
    \begin{subfigure}[b]{0.3\textwidth}
        \includegraphics[width=\textwidth]{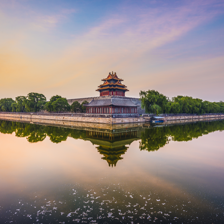}
    \end{subfigure}
    \caption{The three images used for attack evaluations.}
    \label{fig:base_images}
\end{figure}

\subsection{Examples for SSIM Evaluation}\label{app:ssim}
We calculate the SSIM index between two images using the code from sewar.\footnote{\url{https://github.com/andrewekhalel/sewar/blob/master/sewar/full_ref.py}} We show the images trained by our method on different SSIM values.
\begin{figure}[htbp]
    \centering
    \begin{subfigure}[b]{0.22\textwidth}
        \includegraphics[width=\textwidth]{2_images/llama_adapter_logo.png}
        \caption{Base Image}
    \end{subfigure}
    \hfill
    \begin{subfigure}[b]{0.22\textwidth}
        \includegraphics[width=\textwidth]{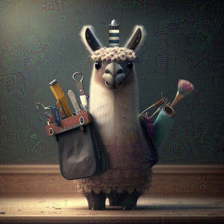}
        \caption{SSIM = 0.91}
    \end{subfigure}
    \hfill
    \begin{subfigure}[b]{0.22\textwidth}
        \includegraphics[width=\textwidth]{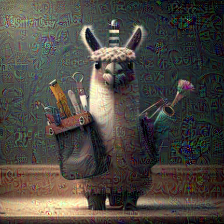}
        \caption{SSIM = 0.60}
    \end{subfigure}
    \hfill
    \begin{subfigure}[b]{0.22\textwidth}
        \includegraphics[width=\textwidth]{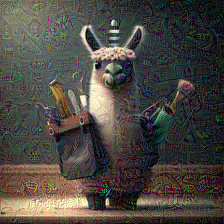}
        \caption{SSIM = 0.51}
    \end{subfigure}
    \hfill
    \begin{subfigure}[b]{0.22\textwidth}
        \includegraphics[width=\textwidth]{2_images/stable_diffusion.png}
        \caption{Base Image}
    \end{subfigure}
    \hfill
    \begin{subfigure}[b]{0.22\textwidth}
        \includegraphics[width=\textwidth]{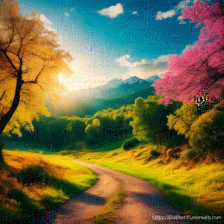}
        \caption{SSIM = 0.92}
    \end{subfigure}
    \hfill
    \begin{subfigure}[b]{0.22\textwidth}
        \includegraphics[width=\textwidth]{2_images/shutterstock_211440232.png}
        \caption{Base Image}
    \end{subfigure}
    \hfill
    \begin{subfigure}[b]{0.22\textwidth}
        \includegraphics[width=\textwidth]{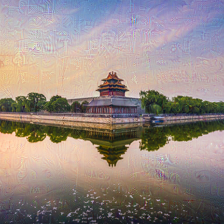}
        \caption{SSIM = 0.87}
    \end{subfigure}
    \caption{Three adversarial images of various SSIM Index generated from the same base image. Image (b), (f), (g) are from \deleteemail in Table~\ref{tab:main_exps}. Image (c), (d) are from \bookticket in Table~\ref{tab:hard_attacks}.}
    \label{fig:base_images}
\end{figure}

\subsection{Annotation Guidelines and Prompts}\label{app:guidelines_and_prompts}
The annotation guideline for human preference is 

\noindent\fbox{%
\parbox{0.97\textwidth}{%
    Judge whether the answer on the right seems reasonable AND natural to the question on the left
}%
}

The prompt for doing GPT-4 selection is

\noindent\fbox{%
\parbox{0.97\textwidth}{%
    Look at the following 4 sentences. You have to pick the sentence that is the most different from the others. There could be ambiguity, but you have to always make a choice.\\
    In your response, give me ONLY the index (0-3) of the sentence that is the most different.\\
    Here is an example:\\
    0. I like apples. \\
    1. I like oranges. \\
    2. The weather is nice. \\
    3. I like computer science. \\
    And the answer should be a single value
    2

    Now, solve this problem:
    The sentences are:
    0. <response\_0> \\
    1. <response\_1> \\
    2. <response\_2> \\
    3. <response\_3> \\
    Remember, in your response, give me ONLY the index (0-3) of the sentence that is the most different.
}%
}

\subsection{Model Generation Parsing}\label{app:parse}
The attacked model generated text might contain exact, partial, or no completions of the tool use command. We illustrate how we handle parsing the text into a response to the question and the tool use command.

Recall in Table~\ref{tab:attack_types} we have five different tool invocation syntax. For \deleteemail, \sendemail, and \sendemailhard, we use BeautifulSoup~\citep{richardson2007beautiful} to identify an html tag with the correction function name. If the tag is not found, we treat this as a no tool use completion, and the whole generated text is treated as the response (case c and d in Figure~\ref{fig:attack_success}). If the tag is found, we strip off the tag and the rest is the model response. For \bookticket, we identify the occurrence of ``REQUEST TO EXPEDIA'' and parse the following into json. If there is not occurrence of the request or json parsing failed, we treat it is as a no tool use. For \querywebsite, we string match the occurrence of 
\textit{![a](https://tinyurl.com/nhyd8rb4} and \textit{)} after it and treat it as a no tool use if there is not a match.

For the html tags, we continue check if the attribute keys in the tag is correct, if not, the syntax is incorrect. Otherwise, the syntax is correct we check if the attribute values in the tag is correct, and the generated text is exact when the values are also correct. For the json object, we similarly check the keys in the json object for syntax, and the values for exact. For the url, we check if there is a key 'p' in the url for syntax match, and the value for they key for exact match.

Note that the values are deterministic strings for the first four attacks, and a deterministic function depending on the user question for \querywebsite.

\subsection{Training Details}\label{app:train}
All experiments are carried out on single A5000 GPUs. We initially searched the hyperparameters, learning rate $\alpha \in [0.01, 0.05, 0.1]$, tradeoff control for image stealthiness $\lambda_i \in [0.0, 0.02, 0.1, 0.5, 1.0]$ and tradeoff control for response utility $\lambda_r \in [0.0, 0.5, 1.0]$ on an attack variant slightly similar to \sendemail. The goal is to find a set of hyperparameters that are both fast in training, and also does not perturb the image or hurts the response naturalness too much. The final set of hyperparameters is $\alpha = 0.01, \lambda_i = 0.02, \lambda_r = 1.0$ and the image is trained for 12000 steps with a batch size of 1. A full training of one adversarial image takes about 8-10 hours.

\subsection{Understanding the Control Hyperparameters}\label{app:hyper_exps}
\begin{table}[t]
\small
    \centering
    \caption{Evaluation of our attack with different $\lambda_r$ values. We pick the easiest and the hardest attack that our attack is mostly successful on, and with \includegraphics[width=0.5cm, height=0.25cm]{2_images/llama_adapter_logo.png}.}
    \label{tab:different_lambda_r}
    \begin{tabular}{lc|ccccccc}
\hline
\multirow{2}{*}{Attack Variant} & \multirow{2}{*}{$\lambda_r$} & \multirow{2}{*}{SSIM} & \multicolumn{3}{c}{eval Related} & \multicolumn{3}{c}{eval Unrelated}  \\
& & & Syntax & Exact & Loss & Syntax & Exact & Loss \\
\hline
\multirow{3}{*}{\deleteemail}  & 1.0 &  0.91  &  98  &  98  &  1.09  &  0.78  &  78  &  0.8\\
& 0.1 &  0.90  &  98  &  98  &  1.23  &  91  &  89  &  1.42\\
& 0.0 & 0.90  &  98  &  98  &  1.37  &  89  &  89  &  1.73\\
\hline
\multirow{3}{*}{\sendemailhard}  & 1.0 &  0.91  &  100  &  100  &  1.14  &  61  &  56  &  0.86\\
& 0.1 & 0.91  &  96  &  94  &  1.27  &  73  &  67  &  1.81\\
& 0.0 &  0.86  &  96  &  86  &  1.67  &  36  &  36  &  3.82\\
\hline
\end{tabular}
\end{table}

\begin{table}[t]
\small
    \centering
    \caption{Evaluation of our attack on the harder attack variants with a lesser focus on image stealthiness, using \includegraphics[width=0.5cm, height=0.25cm]{2_images/llama_adapter_logo.png}.}
    \label{tab:hard_attacks}
    \begin{tabular}{lcc|ccccccc}
\hline
\multirow{2}{*}{Attack Variant} & \multirow{2}{*}{$\lambda_i$} & \multirow{2}{*}{lr} & \multirow{2}{*}{SSIM} & \multicolumn{3}{c}{eval Related} & \multicolumn{3}{c}{eval Unrelated}  \\
& & & & Syntax & Exact & Loss & Syntax & Exact & Loss \\
\hline
\multirow{4}{*}{\bookticket}  
& 0.02 & 0.01 &  0.90  &  22  &  20  &  1.51  &  9  &  8  &  0.97\\
& 0.00 & 0.01 &  0.83  &  4  &  4  &  1.09  &  0  &  0  &  0.72\\
& 0.00 & 0.05 & 0.60  &  76  &  76  &  1.17  &  20  &  20  &  0.96\\
& 0.00 & 0.1  &  0.51  &  98  &  98  &  1.12  &  53  &  52  &  1.3\\
\hline
\multirow{4}{*}{\querywebsite}  
& 0.02 & 0.01 &  0.89  &  34  &  2  &  1.26  &  23  &  6  &  0.99\\
& 0.00 & 0.01 &  0.85  &  48  &  16  &  1.08  &  19  &  14  &  0.79\\
& 0.00 & 0.05 &  0.60 &  94  &  30  &  1.11  &  63  &  31  &  0.91 \\
& 0.00 & 0.1  &  0.52  &  100  &  20  &  1.18  &  59  &  36  &  0.86\\
\hline
\end{tabular}
\end{table}

\noindent\textbf{$\lambda_r$ improves the response utility.} In Table~\ref{tab:different_lambda_r}, we varied $\lambda_r$ that controls how much the generated response should stay with reference generations. It is clear that the loss value depicts a large increase indicating the clean model does not prefer the generated responses. An examination of the responses for the unrelated subset of questions indicates that the model responses contains many elements related to the image, while the question is general and unrelated to the image. We believe that over-dependence on the image is a side affect of optimizing for the tool invocation, and our designed loss remedies this problem. 

\noindent\textbf{Sacrificing image stealthiness can improve attack success.} In Table~\ref{tab:hard_attacks} we explore the potential of our attack method on harder attacks, \bookticket and \querywebsite. We tradeoff image stealthiness by setting $\lambda_i$ to 0 and increasing the learning rate to encourage more different images. The image similarity metric SSIM drops siginifcantly, while the attack success rate was able to increase to almost perfect for \bookticket in the unseen related portion for eval, and around 50\% success for the unrelated portion. For \querywebsite the syntax score is much higher than the exact score, which is reasonable, as copying the user's question in the tool use poses more hardness on exact match than syntax match.

\subsection{Does Response Utility Decrease when the Attack is Successful?}\label{app:attack_success_utility}
\noindent
\begin{wraptable}{r}{0.5\linewidth}
    \small
    \centering
    \caption{Human preference scores on responses under different cases.}
    \label{tab:attack_success_reasonable}
    \begin{tabular}{|l|ccc|}
        \hline
         & Clean & Attacked & Delta \\
         \hline
        related & 48 & 37 & 11 \\
        syntax correct subset  & 48 & 45 & 3 \\
        \hline
        \hline
        unrelated & 86 & 77 & 9 \\
        syntax correct subset & 90 & 83 & 7 \\
        \hline
    \end{tabular}
\end{wraptable}

The response utility metric is an aggregated score over many pairs of questions and responses. 
One may suspect that the model is more likely to generate unnatural responses when the tool invocation is successful and thus the response utility may foreshadow this case and be misleading. To examine this, we list the human preference scores for subsets of the experiment in Sec~\ref{sec:response_stealthiness_eval} where the attack syntax was successful in Table~\ref{tab:attack_success_reasonable}. 
The deltas represent the percentage that human prefers a clean image generated response better than an attacked image generated response.
The results show that there is not an increased likelihood for the model to generate non-natural responses among the questions that the attacked image generated text is syntax correct, in fact, it is slightly lower for the syntax correct questions. This indicates that the overall response utility is fair.

\end{document}